# Structural and magnetic properties of the $(Bi_{2-x}Pr_x)Ru_2O_7$ pyrochlore solid solution (0 ≤ x ≤ 2)

S. Zouari[a,b], R. Ballou[b] , A. Cheikhrouhou[a], P. Strobel[b]

[a]Laboratoire de Physique des Matériaux, Faculté des Sciences de Sfax, BP 763, 3038 Sfax, Tunisia

[b]Institut Néel, CNRS, B. P. 166, 38042 Grenoble, Cedex 9 France.

**Abstract**

We report a detailed study of structural and magnetic properties of the pyrochlore-type $(Bi_{2-x}Pr_x)Ru_2O_7$ series. Eleven compositions with 0 ≤ x ≤ 2 were prepared by solid state reaction at 1050°C and found to form a solid solution in the whole range of compositions. Structural refinements from X-ray powder diffraction data by the Rietveld method show that the main variation concerns the ruthenium site. Ru-O distances increase and the Ru-O-Ru angle (bridging corner-sharing Ru-O octahedra) decreases with increasing praseodymium content, in agreement with the metal-semiconductor transition observed previously in this system. Magnetic measurements show that the metallic Pauli paramagnetism of Ru in $Bi_2Ru_2O_7$ evolves to a magnetism of localized low spin moments coexisting with the magnetism of $Pr^{3+}$ cations moments. Strong magnetic correlations are present at high Pr content, as illustrated by a Néel temperature of -224 K for $Pr_2Ru_2O_7$. However no evidence of a magnetic ordering was found, suggesting that the end compound $Pr_2Ru_2O_7$ of the series might stabilize a spin liquid phase.

_Corresponding author:_  P. STROBEL

Institut Néel, CNRS, case F, BP 166, F-38042, Grenoble Cedex 9, France

phone: 33+ 4 76 88 79 40

fax: 33+ 4 76 88 10 38

e-mail:  pierre.strobel@grenoble.cnrs.fr

## 1. Introduction

Oxides with the pyrochlore–type structure have attracted great attention recently because of their unique cationic sublattice of corner-sharing tetrahedra, providing magnetic systems with geometrical frustration and interesting physical properties at low temperatures [1]. The pyrochlore structure is cubic, space group Fd-3m (n°227), with eight $A_2B_2O_7$ formula units per unit cell [2, 3]. The most common charge distribution are $A_2^{2+}B_2^{5+}O_7^{2-}$ and $A_2^{3+}B_2^{4+}O_7^{2-}$.

Pyrochlores-type ruthenates $A^{3+}_2Ru^{4+}_2O_7$ where the A cation is a trivalent rare earth raised a particular interest. According to previous studies [4-6], these compounds exhibit magnetic transitions for A = Pr, Nd, Sm, Eu and Y at 165, 150, 135, 120 and 145 K, respectively, suggesting contributions to the magnetism from both the trivalent rare earth and $Ru^{4+}$. For $Nd_2Ru_2O_7$, $Sm_2Ru_2O_7$, and $Eu_2Ru_2O_7$, evidence for a spin glass state below this temperature has been reported, as well as coexistence of a weak ferromagnetic state with the spin glass state below 20 K [5].

The solid solutions $Bi_{2-x}Ln_xRu_2O_7$ (Ln = Pr – Lu) have been investigated previously for some lanthanides [7-9]. The unsubstituted compound $Bi_2Ru_2O_7$ is metallic and $Bi_{2-x}Ln_xRu_2O_7$ solid solutions have been shown to cross a metal-semiconductor transition on increasing the rare earth content x [9]. The lattice parameter increases with x for the light rare earth cations (Pr, Nd) and decreases with x for the heavy rare earth cations (Sm, Dy). The ruthenium and oxygen networks show similar distortions, which would induce the electronic transitions, independently of the nature of the rare earth element. Magnetic properties were reported only for Ln = Nd [9]. We also note that in the previous study of $Bi_{2-x}Ln_xRu_2O_7$ systems [9], no XRD diagrams were shown, and that the interatomic distances obtained for unsubstituted $Bi_2Ru_2O_7$ disagree with those reported in other studies [10-12]. The lack of reports of magnetic properties and the inconsistencies in interatomic distances (and certainly angles) in the previous studies of $Bi_{2-x}Pr_xRu_2O_7$ prompted us to undertake a detailed crystal chemical and magnetic study of this system, the results of which are presented in this paper.

## 2. Experimental

Polycrystalline $(Bi_{2-x}Pr_x)Ru_2O_7$ samples were prepared using the standard ceramic processing technique by mixing $Pr_6O_{11}$, $Bi_2O_3$ and $RuO_2$, 99.9 % pure, in appropriate proportions. The precursors were thoroughly mixed in an agate mortar, the mixture was pelletized and heated repeatedly at 1050°C in a dry argon flow for 72h, with intermediate grinding. This procedure was adopted following attempts in air giving an almost systematic formation of the impurity phase $Bi_3Ru_3O_{11}$.

Samples were characterized by powder X-ray diffraction (XRD) at room temperature, using a Bruker D8 diffractometer equipped with a Kevex Si(Li) solid detector and Cu K$a$ radiation. Diffraction patterns were recorded in transmission mode with 0.02° steps and counting time 50-70 seconds per step. Crystal structures were refined from powder X-ray data by the Rietveld method, using the Fullprof program [13]. The background was interpolated between manually selected points. A pseudo-Voigt profile was used, including the Caglioti function variables U, V, W, the mixing parameter η, and one or two asymmetry parameters. Other refine variable were the scale factor, zero shift, atomic coordinates and isotropic atomic displacement parameters.

In order to check for possible oxygen non-stoichiometry, several samples were submitted to thermogravimetry in oxygen up to 800°C, using a Setaram TAG-1600 thermo-analyzer. Magnetic isotherms were measured using purpose-built axial extraction magnetometers in the temperature range 2–300 K under applied fields up to 10 T, and on $Pr_2Ru_2O_7$ in the temperature range 300–800 K under applied fields up to 5 T.

## 3. Results and discussion

### 3.1. Crystal chemistry

Typical X-ray diffraction patterns of $(Bi_{2-x}Pr_x)Ru_2O_7$ compounds ($0 \leq x \leq 2$) are shown in Fig.1. All samples crystallize in the cubic system with Fd-3m space group and form a continuous solid solution. Weak impurities – only noticeable in high signal-to-noise acquisitions such as those displayed in Fig.1 – are present for x = 1.2 and for x = 2. They correspond to metallic ruthenium and to $Pr_{11}O_{20}$, respectively. The former may arise from the use of an argon atmosphere in the synthesis to prevent the formation of $Bi_3Ru_3O_{11}$; the reason why it occurs for $1.0 \le x \le 1.6$ is not known. In all cases the impurities represent less than 0.5 % in weight and have a negligible effect on structural refinements.

The cubic lattice parameter $a$ increases monotonously with praseodymium content x, in agreement with previous data (see Fig. 2). The structure was refined in the pyrochlore structure model, i.e. with A = (Bi,Pr) atoms on 16d site (½ ½ ½), ruthenium on 16c (0 0 0), O1 on 48f (x 1/8 1/8), and O2 on 8b site (3/8 3/8 3/8). The results of Rietveld refinements are summarized in Table 1. Standard deviations have been corrected for serial correlation effects [14]. Significant refinements are displayed in Fig. 3. The Bi/Pr atoms on the pyrochlore A site systematically exhibit considerably higher atomic displacement parameters (ADP) than ruthenium atoms. This is not surprising, considering that the A site coordination is much less regular : it consists of a heavily distorted cube with two sets of distances to O1 (6 neighbours) and O2 (2 neighbours) [2]; it must also accommodate the bismuth lone pair, an additional cause of site distortion.

Oxygen vacancies are known to be easily accommodated on the O2 site in the pyrochlore structure [2]; however this is impossible to determine accurately from X-ray diffraction data in the presence of heavy atoms such as bismuth and lanthanides; moreover the O2 site involves only 1/7th of the oxygen atoms. Using neutron diffraction, oxygen vacancy levels up to 0.20 per formula unit in $Bi_2Ru_2O_{7-\delta}$ were reported after appropriate treatments in reducing atmosphere and yielded an increase in cell parameter to $a = 10.299$ Å [14]. More recently, Avdeev et al. reported that substituted $(Bi_{2-x}Ln_x)Ru_2O_7$ phases do not contain appreciable oxygen non-stoichiometry [12]. We checked the oxygen content using thermogravimetry in oxygen. All our measurements on substituted $(Bi_{2-x}Pr_x)Ru_2O_7$ phases yielded very small mass variations, corresponding to oxygen content variations lower than 0.02 per formula unit. From these measurements, and from the value of cell parameter for $Bi_2Ru_2O_7$ (10.2925 Å, compared to 10.299 Å for $Bi_2Ru_2O_{6.80}$ [15]), we conclude that the pyrochlore compounds under investigation here have a negligible oxygen vacancy level.

Significant interatomic distances and bonding angles are given in Table II. Most variations are rather small (close to 0.8 % on the cell parameter and (Bi/Pr)-O distances between x = 0 and x = 2); they are consistent with the small difference between the ionic radii of $Bi^{3+}$ and $Pr^{3+}$ (1.17 and 1.13 Å for 8-fold coordination, respectively [16]). All atoms but O1 occupy fixed positions in the structure. The effect of a variation in x(O1) is as follows [2, 17]: an increase in x(O1) brings the O1 atom closer to atom A (Bi,Pr) and farther from atom B (Ru). In addition, a larger x(O1) value makes the B-(O1)$_6$ octahedron less regular while it decreases the distortion of the A-O$_8$ cube. As a result, the Ru-O distance increases with praseodymium content (see Fig.4a), whereas the antagonistic effects of the cell parameter $a$ and x(O1) on the A-O1 distance explain the peculiar evolution of the A-O1 bond length (Fig.4b). This distance is almost constant up to x ≈ 1, then decreases with increasing praseodymium content due to the dominant effect of x(O1). On the contrary, the A-O2 distance, involving only atoms in fixed positions, exhibits a monotonous increase wth increasing x(Pr) (Fig. 4c). Table 1 includes literature data for Bi$_2$Ru$_2$O$_7$ and Pr$_2$Ru$_2$O$_7$; it can be noted that the (Bi/Pr)-O1 and Ru-O1 distances obtained here are in much better agreement with those determined from neutron data [10] than those reported in ref.9, where x(O1) is likely to be in error.

In the pyrochlore structure, Ru-O1 octahedra form zig-zag chains characterized by the Ru-O-Ru bridging angle. Several authors have tentatively correlated the occurrence of a metal-semiconductor transition with the Ru-O bond length and the Ru-O-Ru angle [9, 16]. In this respect, it is interesting to note that the ruthenium site size, which is driven in the same direction by the simultaneous increase in both $a$ and x(O1), is more affected by the Bi/Pr substitution than the Bi/Pr site itself : the Ru-O1 distance increases by 1.8 % between x = 0 and x = 2, compared to 0.8 % for Bi/Pr-O distances. This indirect effect of the A cation size has been confirmed by electronic structure calculations [18]. In addition, the distortion in the RuO6 octahedron increases with x(O1), i.e. with praseodymium content : the O-Ru-O angle shift from 90° increases significantly from x = 0 (84.9/95.1°) to x = 2 (83.1/86.9°), as shown in Fig. 4d. Similarly, the bridging Ru-O-Ru angle (see Fig. 4e) decreases with increasing x and indeed crosses the boundary recently predicted for the occurrence of a metal-non metal transition [11].

### 3.2. Magnetic properties

Fig. 5 shows a few magnetic isotherms M(H, T) measured at different temperatures T on compounds in the Bi-rich (x = 0.6) and Pr-rich (x = 1.8) sides of the (Bi$_{2-x}$Pr$_x$)Ru$_2$O$_7$ solid solution. These are linear in the available range of the applied magnetic field for all temperatures above 40 K. At lower temperatures the magnetic isotherms are bended downwards in the Pr-rich compound but do not saturate up to 10 T. They can be fitted to the function M(H, T) = χ(T) H + χ$^{(3)}$(T) H$^3$. On increasing Pr content up to pure Pr$_2$Ru$_2$O$_7$ (x = 2), the non-linear magnetic susceptibility $\chi^{(3)}$ gets larger. Nevertheless, no saturation is reached up to 10 T, which might indicate strong crystalline electric field effects or a possible magnetic order without a spontaneous moment. On decreasing Pr content, a decrease of the non linear magnetic susceptibility $\chi^{(3)}$ is observed.

The thermal variation of the linear magnetic susceptibility $\chi$ extracted from the magnetic isotherms measured in the $Bi_2Ru_2O_7$ compound is shown in Fig. 6. Except a slight increase at the lowest temperatures, which might be of extrinsic origin, it is almost temperature-independent and consistent with a Pauli paramagnetism of metallic electrons [9]. Note that the magnetic isotherms measured in this compound are linear down to the lowest temperatures within the accuracy of the measurements (see inset of Fig. 5).

The thermal variation of the inverse linear magnetic susceptibility $\chi^{-1}$ extracted from the magnetic isotherms measured in $Pr_2Ru_2O_7$ is shown in Fig. 7. A Curie-Weiss behavior is evidenced at high temperatures. A fit to the function $\chi^{-1}(T) = (T-\theta)/C$ leads to a Curie constant C = 0.106 T / ($\mu$B / f.u. K) and a Néel temperature $\theta = -224$ K. The value of C corresponds to a squared effective moment $\mu_{eff}^2 = 42.1 \ \mu_B^2$ /f.u. On subtracting to it the quantity $2[\mu_{eff}(Pr^{3+})]^2 = 2[3.57771]^2$, where $\mu_{eff}(Pr^{3+})$ is the effective moment of the free $Pr^{3+}$ ion, we find $\mu_{eff}^2 - 2[\mu_{eff}(Pr^{3+})]^2 = 16.5 \ \mu_B^2$. This defines an effective moment $\mu_{eff}(Ru^{4+}) = 2.87 \ \mu_B^2$ that can be be ascribed to the $Ru^{4+}$ 4d electrons. This value of $\mu_{eff}(Ru^{4+})$ is very close to $2(S(S+1))^{1/2}$ with S = 1, suggesting that the $Ru^{4+}$ 4d electrons are in the low spin orbital triplet $(t_{2g})^4$ crystalline electric field state [19]. The negative and strong value of $\theta$ reveals dominant antiferromagnetic exchange interactions. Nevertheless, the experimentally determined $\chi^{-1}(T)$ shows neither an anomaly at any specific temperature that would reveal the occurrence of a Néel antiferromagnetic phase nor any magnetic irreversibility that could signal any spin glass phase. A non linear monotonous decrease with the temperature is instead observed that only suggest magnetic correlations without a diverging correlation length, namely a possible spin liquid phase.

## 4. Conclusions

This investigation of the pyrochlore-type $(Bi_{2-x}Pr_x)Ru_2O_7$ ($0 \leq x \leq 2$) solid solution gives a detailed picture of the structural variations in such a system exhibiting a metal-semconductor transition. The structural evolution is explained by the combined effect of the variations in cell parameter and in the x(O1) atomic parameter. The net results are a larger increase of the ruthenium site size compared to the (Bi,Pr) one, and an stronger distortion of the Ru-O$_6$ octahedron. The evolution of distances and angles is consistent with the transition from a metallic state to a semiconducting one with increasing rare earth concentration. For small praseodymium content, the magnetic properties of the compounds are answerable to the addition of an itinerant electron paramagnetism associated with metallic ruthenium 4d electrons and a localized magnetism associated with Pr$^{3+}$ cations randomly distributed on a lattice of corner-sharing tetrahedra. Pr-Pr magnetic correlations take place as Pr content is increased and before percolation threshold is reached. Evidence is provided by the fact that the magnetic susceptibility is not additive in the Pr content even when x ≤ 0.6. For large Pr content, the magnetic properties are answerable to a combination of a magnetism of localized Ru$^{4+}$ 4d low spin moments and of localized Pr$^{3+}$ moments both in a strong crystalline electric field, experiencing strong antiferromagnetic exchange interactions and being distributed on geometrically frustrating lattice of corner-sharing tetrahedral. No condensation into a Néel magnetic phase nor into a spin glass phase is evidenced, but strong magnetic correlations are revealed suggesting a possible spin liquid phase. A more detailed investigation of this low temperature phase will be considered in the near future.

*Acknowledgments*

The authors gratefully acknowledge Region Rhône-Alpes for financial support of S. Zouari (MIRA Program).

**Figure Captions**

Fig. 1. X-ray diffraction patterns of $(Bi_{2-x}Pr_x)Ru_2O_7$ compounds for various Pr contents x. Pyrochlore reflections indexation is indicated.

Fig. 2. Evolution of the cell parameter with x in the $(Bi_{2-x}Pr_x)Ru_2O_7$ series. Open symbols: ref.9, closed symbols: this work (e.s.d.'s are smaller than the symbol size).

Fig. 3. Observed (points) and calculated (continuous line) X-ray diffraction patterns of $(Bi_{2-x}Pr_x)Ru_2O_7$ with x = 0.8 and 1.8 (values of x not included in Fig.1 were purposedly chosen). The lower part shows the difference $I_{obs}$-$I_{calc}$.

Fig. 4. Evolution of interatomic distances and angles with x in the $A_2Ru_2O_7$ series ($A_2$ = $Bi_{2-x}Pr_x$). (a) Ru-O1 distance, (b) A-O1 distance, (c) A-O2 distance, (d) O-Ru-O angle, (e) Ru-O1-Ru bridging angle. In (c), the e.s.d.'s are smaller than the size of the symbols. The lines are a guide to the eye.

Fig. 5. Magnetic isotherms measured at various temperatures on $(Bi_{1.4}Pr_{0.6})Ru_2O_7$ and $(Bi_{0.2}Pr_{1.8})Ru_2O_7$.

Fig. 6. Thermal variation of the magnetic susceptibility in $Bi_2Ru_2O_7$. The inset shows the magnetic isotherm measured at 5 K.

Fig. 7. Thermal variation of the inverse linear magnetic susceptibility in $Pr_2Ru_2O_7$.

Table I. Summary of Rietveld refinements results in the series $Bi_{2-x}Pr_xRu_2O_7$; all distances in Å.

| $x$ | $a$ (Å) | $x(O1)$ | $B(Bi/Pr)$ | $B(Ru)$ | $B(O1)$ | $N-P+C$ | $R_{wp}$ | $\chi^2$ | $R_{exp}$ | $R_{Bragg}$ |
|---|---|---|---|---|---|---|---|---|---|---|
| 0 | 10.2925(3) | 0.326(2) | 0.68(9) | 0.25(10) | 1 † | 4222 | 2.94 | 2.37 | 1.91 | 2.62 |
| *Ref.* | *10.2934* [9] | *0.314(6)* | | | | | | | | |
| *values* | *10.2957* [10,17] | *0.327* | | | | | | | | |
| | *10.2897* [18] | *0.3265(6)* | | | | | | | | |
| 0.2 | 10.3048(8) | 0.327(4) | 0.70(13) | 0.18(16) | 1 † | 4276 | 11.0 | 2.67 | 6.73 | 7.37 |
| 0.4 | 10.3143(5) | 0.323(3) | 0.85(9) | 0.16(11) | 1 † | 4060 | 7.21 | 2.07 | 5.02 | 4.68 |
| 0.6 | 10.3234(3) | 0.326(2) | 1.18(10) | 0.30(12) | 1 † | 4961 | 8.22 | 1.38 | 6.98 | 3.99 |
| 0.8 | 10.3295(2) | 0.327(2) | 1.23(7) | 0.42(9) | 1.3(6) | 3510 | 10.6 | 5.39 | 4.56 | 5.39 |
| 1.0 | 10.3416(3) | 0.326(3) | 1.10(10) | 0.25(12) | 1.0(8) | 3505 | 14.3 | 9.13 | 4.73 | 1.98 |
| 1.2 | 10.3434(3) | 0.326(3) | 1.01(11) | 0.28(13) | 0.9(8) | 3529 | 14.5 | 9.34 | 4.74 | 1.76 |
| 1.4 | 10.3540(2) | 0.329(3) | 1.04(10) | 0.48(12) | 1.1(8) | 3475 | 13.6 | 8.53 | 4.67 | 1.30 |
| 1.6 | 10.3620(2) | 0.330(2) | 0.91(6) | 0.51(8) | 1.0(5) | 3425 | 8.47 | 2.85 | 5.02 | 0.98 |
| 1.8 | 10.3680(2) | 0.330(2) | 0.75(8) | 0.52(10) | 0.5(6) | 3425 | 9.60 | 3.40 | 5.21 | 1.24 |
| 2.0 | 10.3735(4) | 0.331(2) | 0.54(9) | 0.32(11) | 0.6(6) | 3423 | 8.57 | 2.55 | 5.36 | 1.54 |
| | *10.3714* [9] | *0.326(4)* | | | | | | | | |

Table II. Selected bond lengths (Å) and angles (°) in the $Bi_{2-x}Pr_xRu_2O_7$ series.

| X | A-O1 | A-O2 | Ru-O | O1-Ru-O1 | Ru-O1-Ru | Bi-O1-Bi |
|---|---|---|---|---|---|---|
| 0 | 2.555(13) | 2.228(1) | 1.979(11) | 84.9(4) | 133.6(3) | 90.8(4) |
| *Ref. 9* | *2.63* | *2.228* | *1.94* | | | |
| *Ref. 10* | *2.550* | *2.229* | *1.983* | | | |
| 0.2 | 2.551(21) | 2.231(1) | 1.986(20) | 84.5(7) | 133.1(6) | 91.1(6) |
| 0.4 | 2.57(2) | 2.235(1) | 1.975(25) | 85.5(7) | 134.2(6) | 90.4(6) |
| 0.6 | 2.563(15) | 2.235(1) | 1.985(11) | 84.9(4) | 133.6(3) | 90.8(4) |
| 0.8 | 2.558(13) | 2.236(1) | 1.990(13) | 84.6(5) | 133.1(4) | 91.1(5) |
| 1.0 | 2.562(16) | 2.239(1) | 1.992(15) | 84.6(5) | 133.2(5) | 91.1(5) |
| 1.2 | 2.562(20) | 2.239(1) | 1.991(20) | 84.6(7) | 133.3(5) | 91.0(5) |
| 1.4 | 2.547(15) | 2.242(1) | 2.004(15) | 83.7(7) | 131.9(5) | 91.9(5) |
| 1.6 | 2.544(11) | 2.243(1) | 2.009(10) | 83.4(4) | 131.5(3) | 92.1(3) |
| 1.8 | 2.541(12) | 2.245(1) | 2.013(10) | 83.3(4) | 131.2(3) | 92.3(3) |
| 2.0 | 2.540(10) | 2.246(1) | 2.015(9) | 83.1(4) | 131.0(3) | 92.4(3) |
| *Ref.9* | *2.58* | *2.246* | *1.993* | | | |

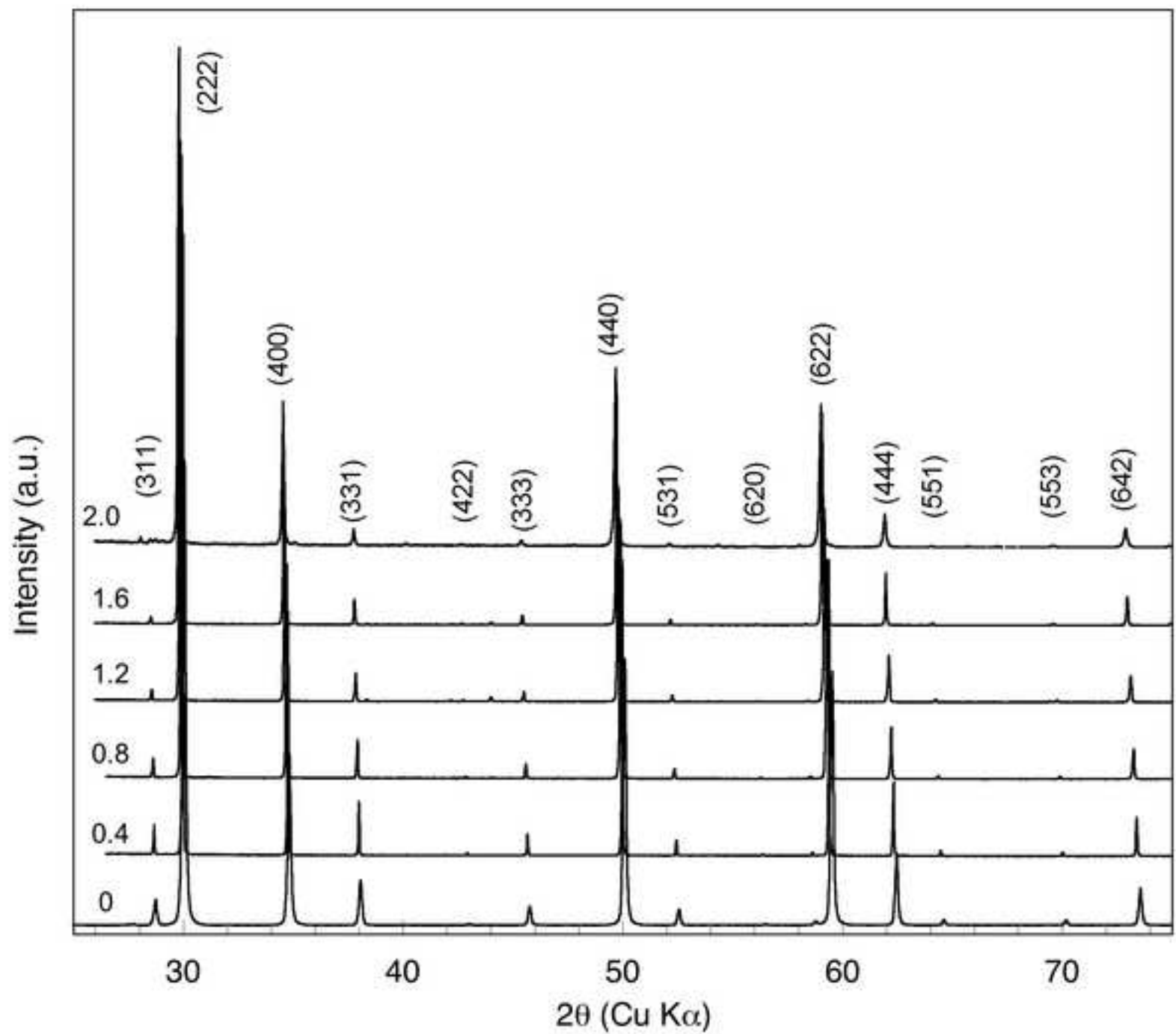

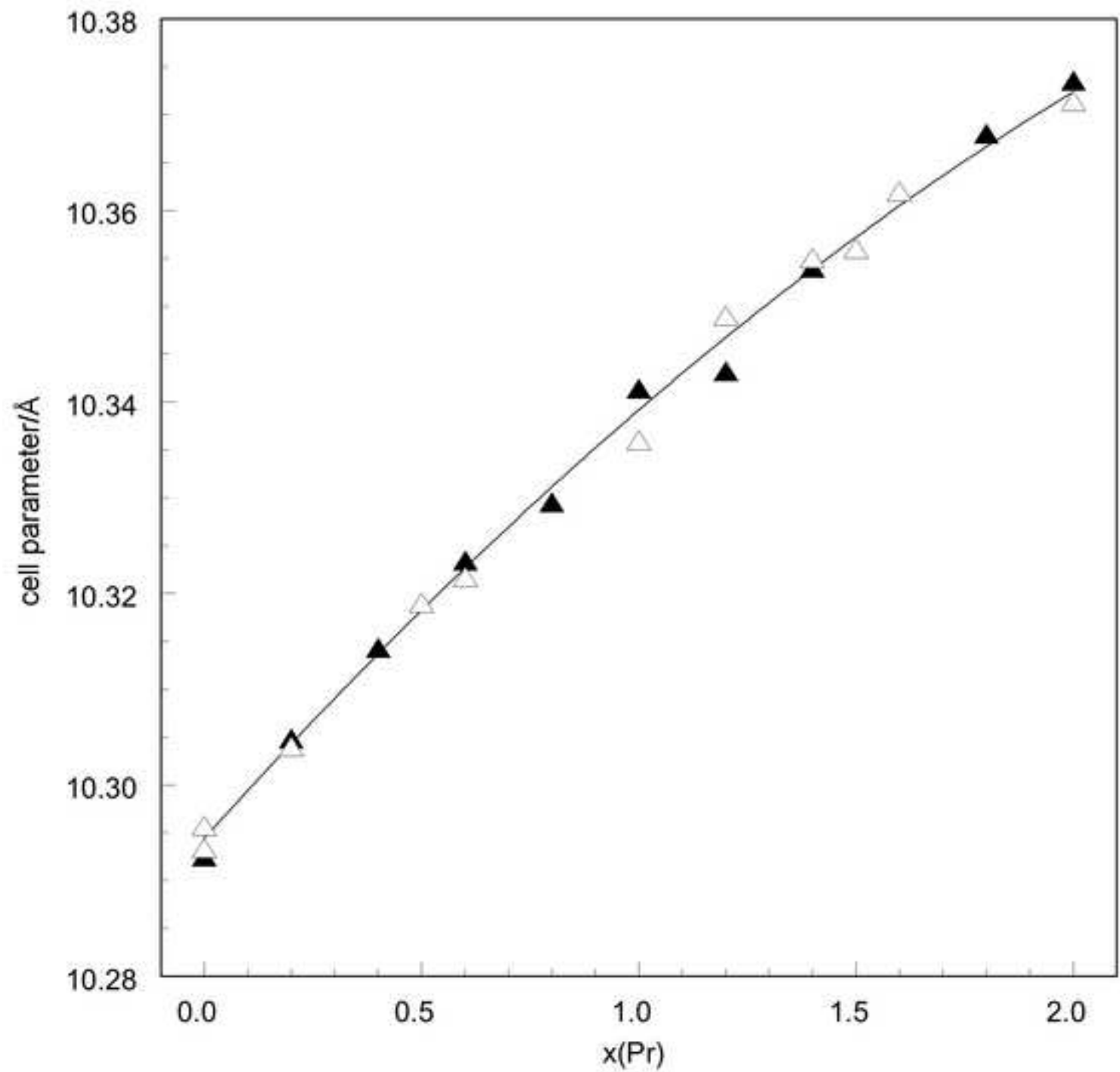

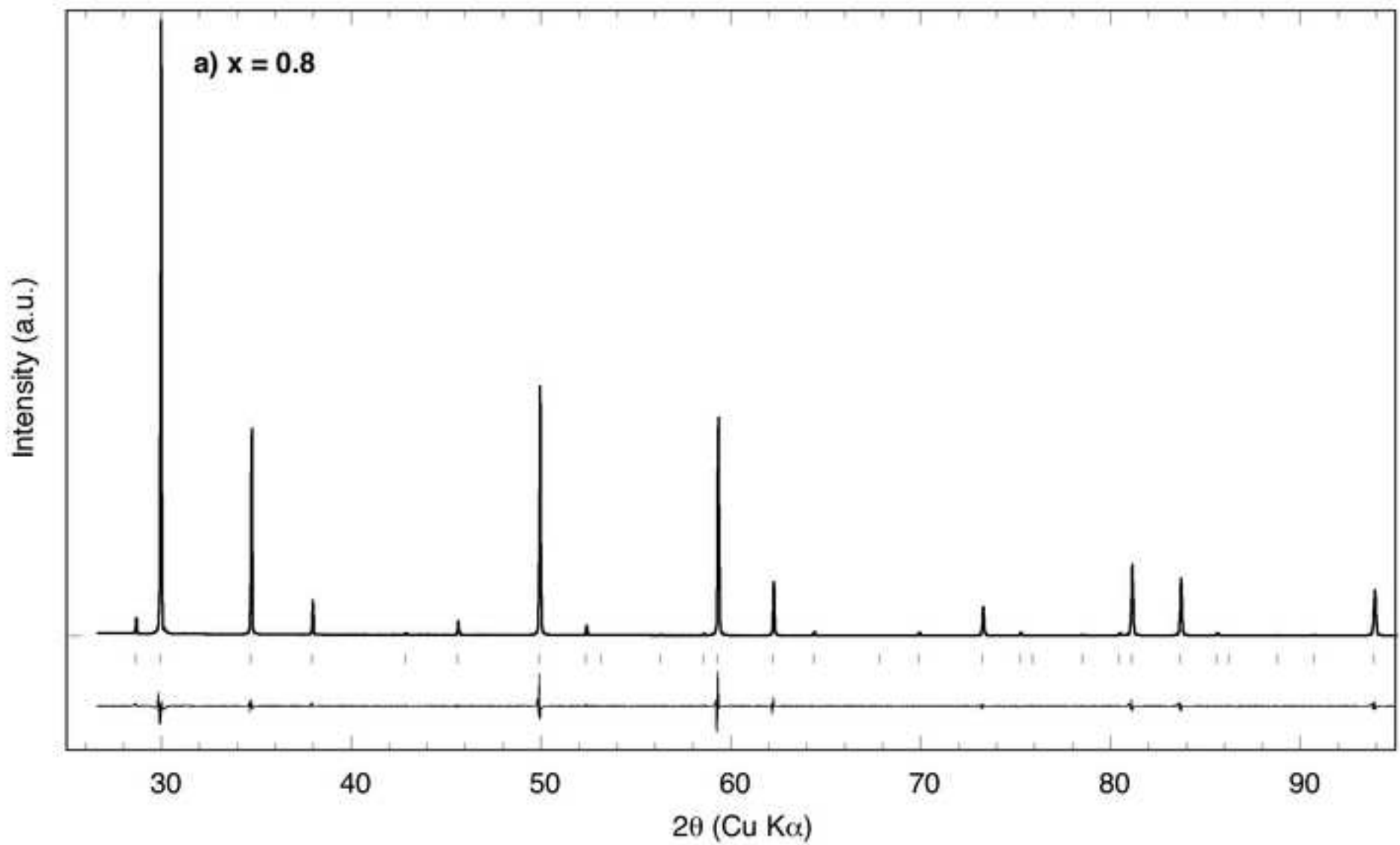

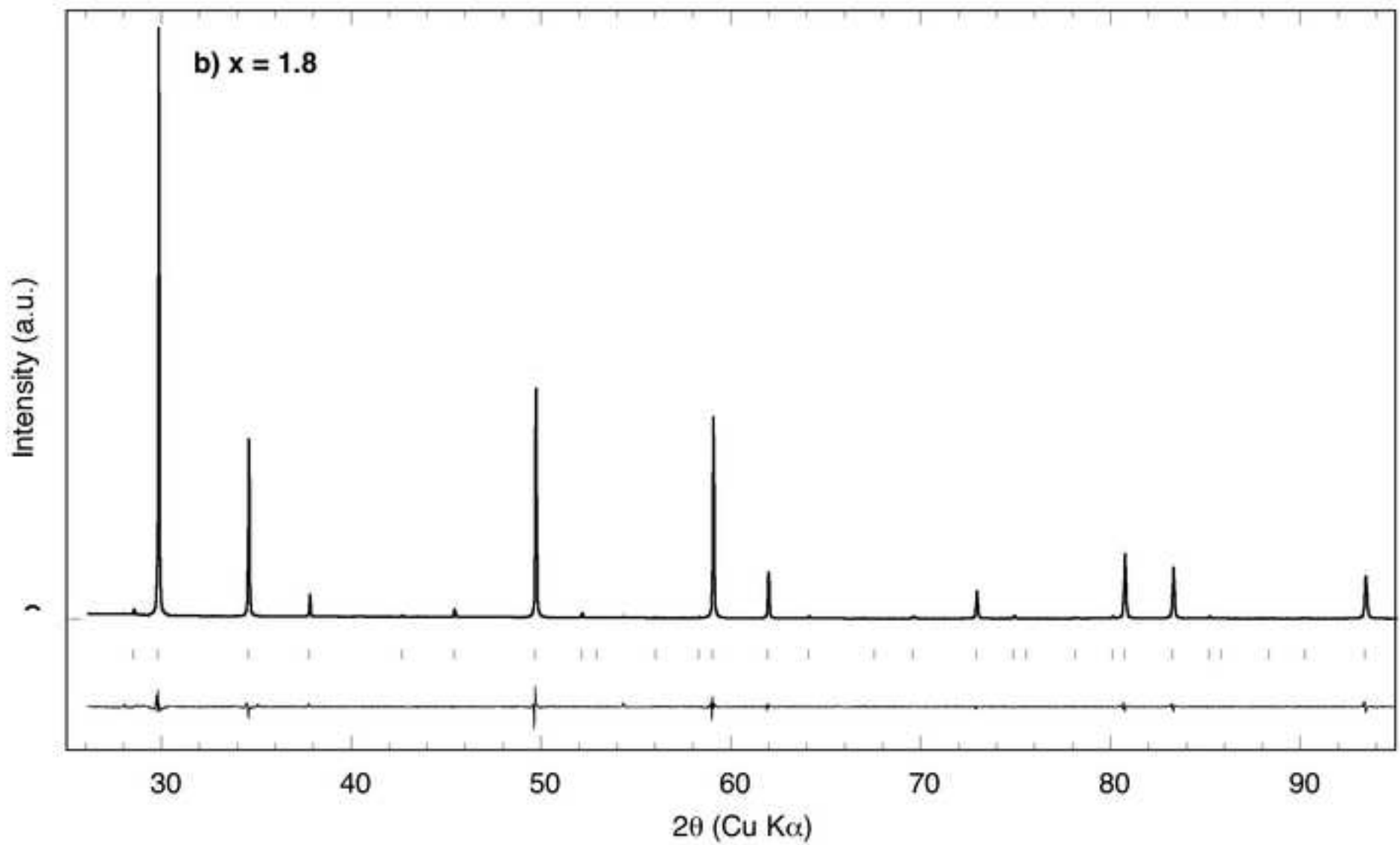

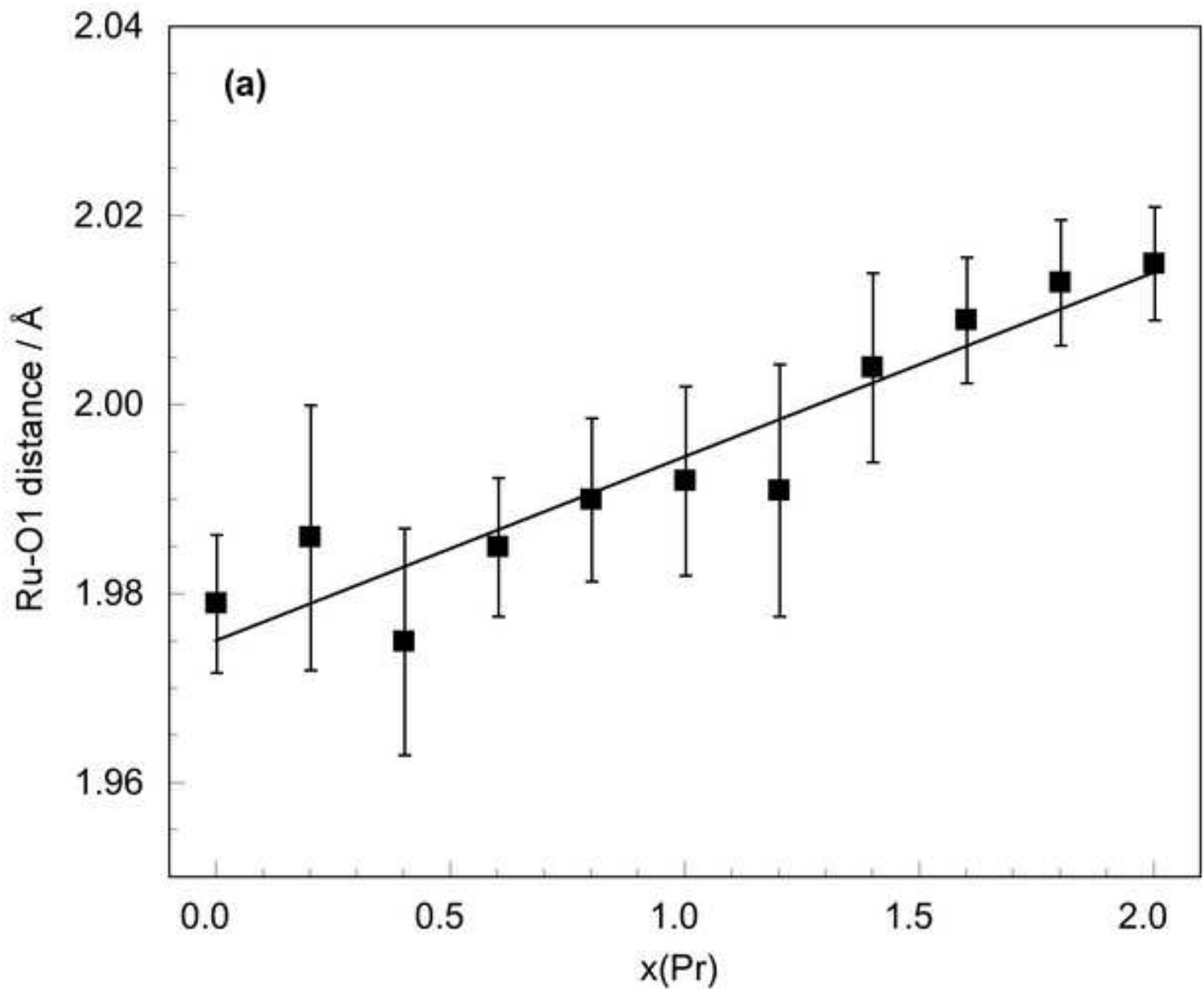

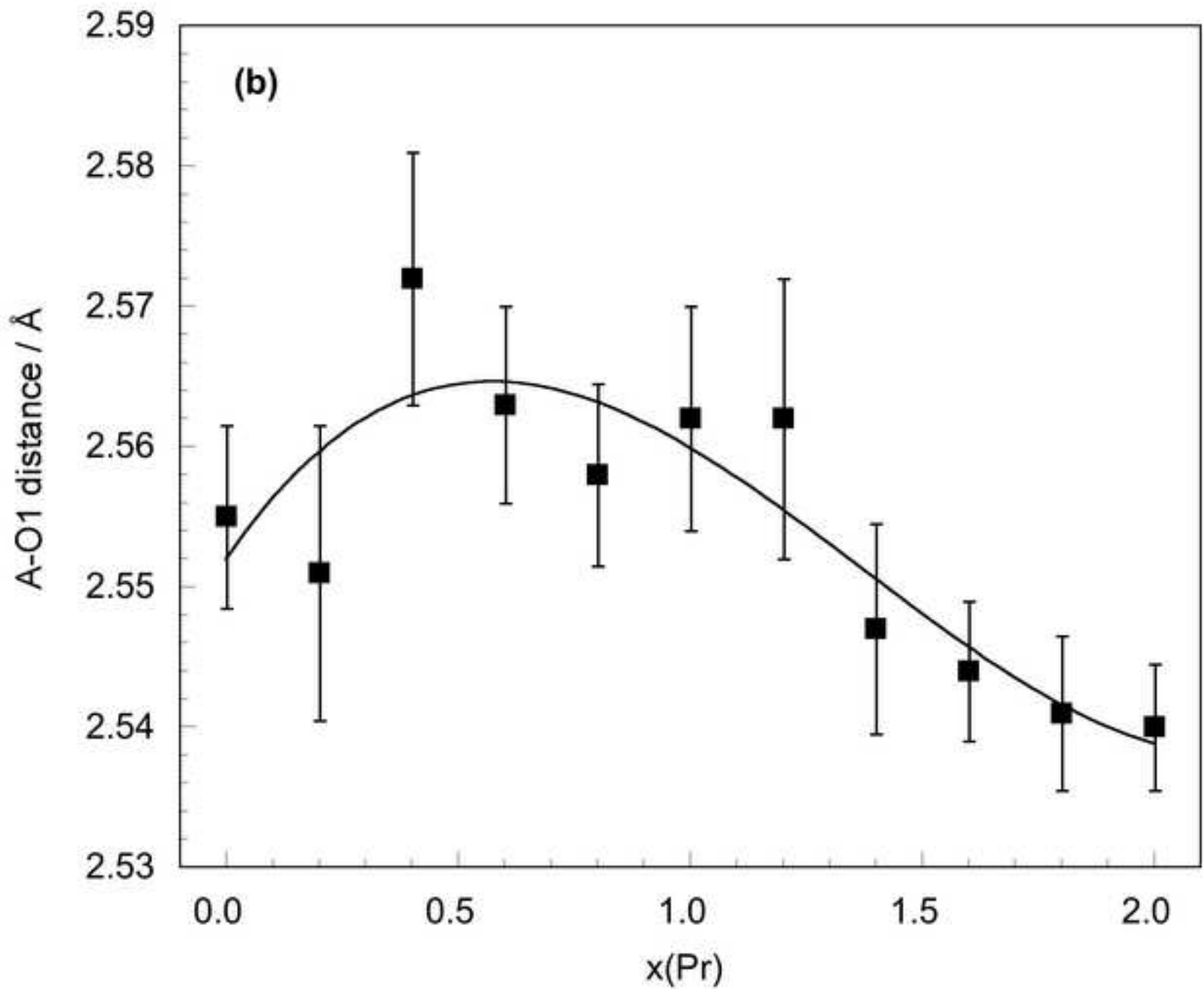

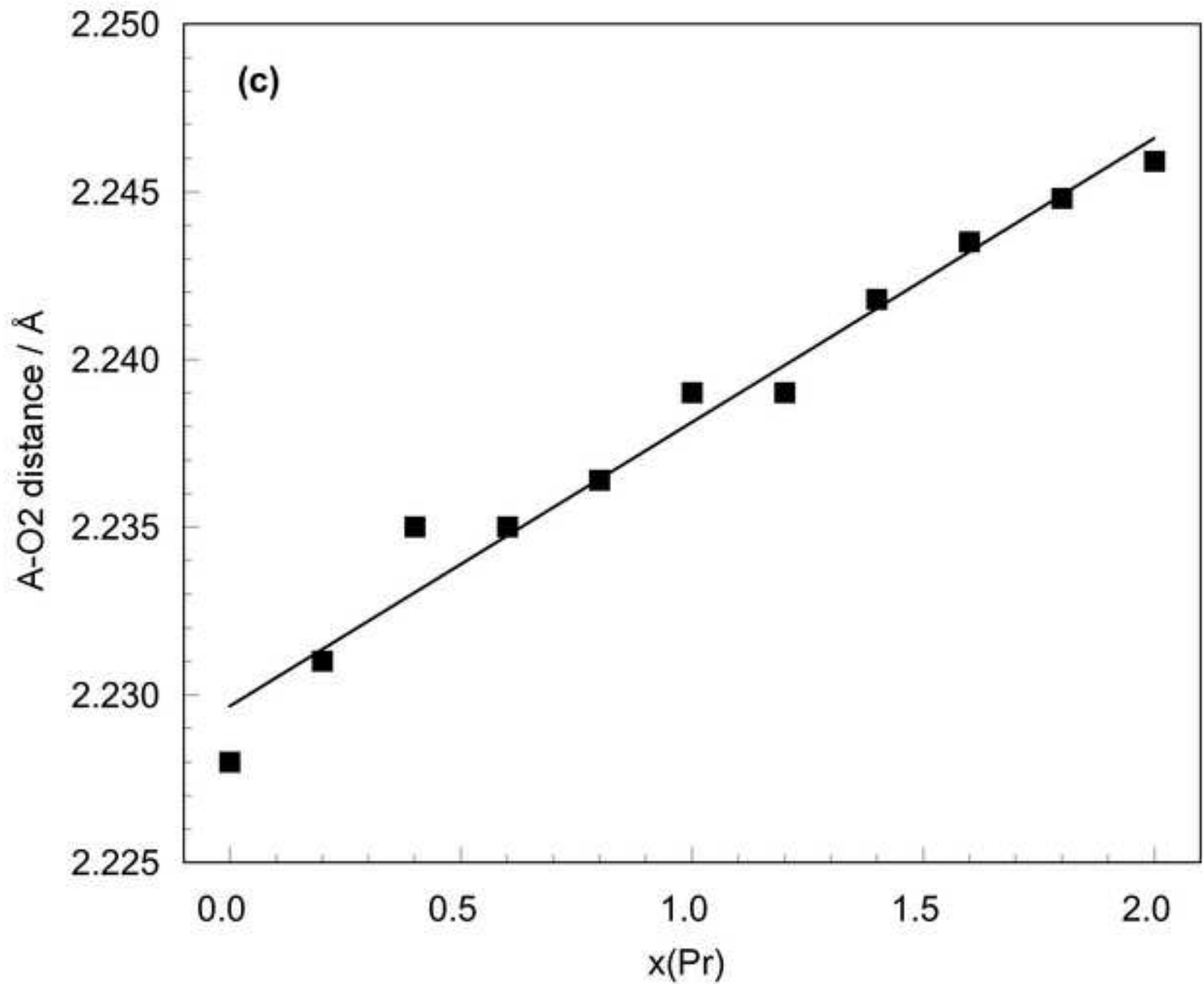

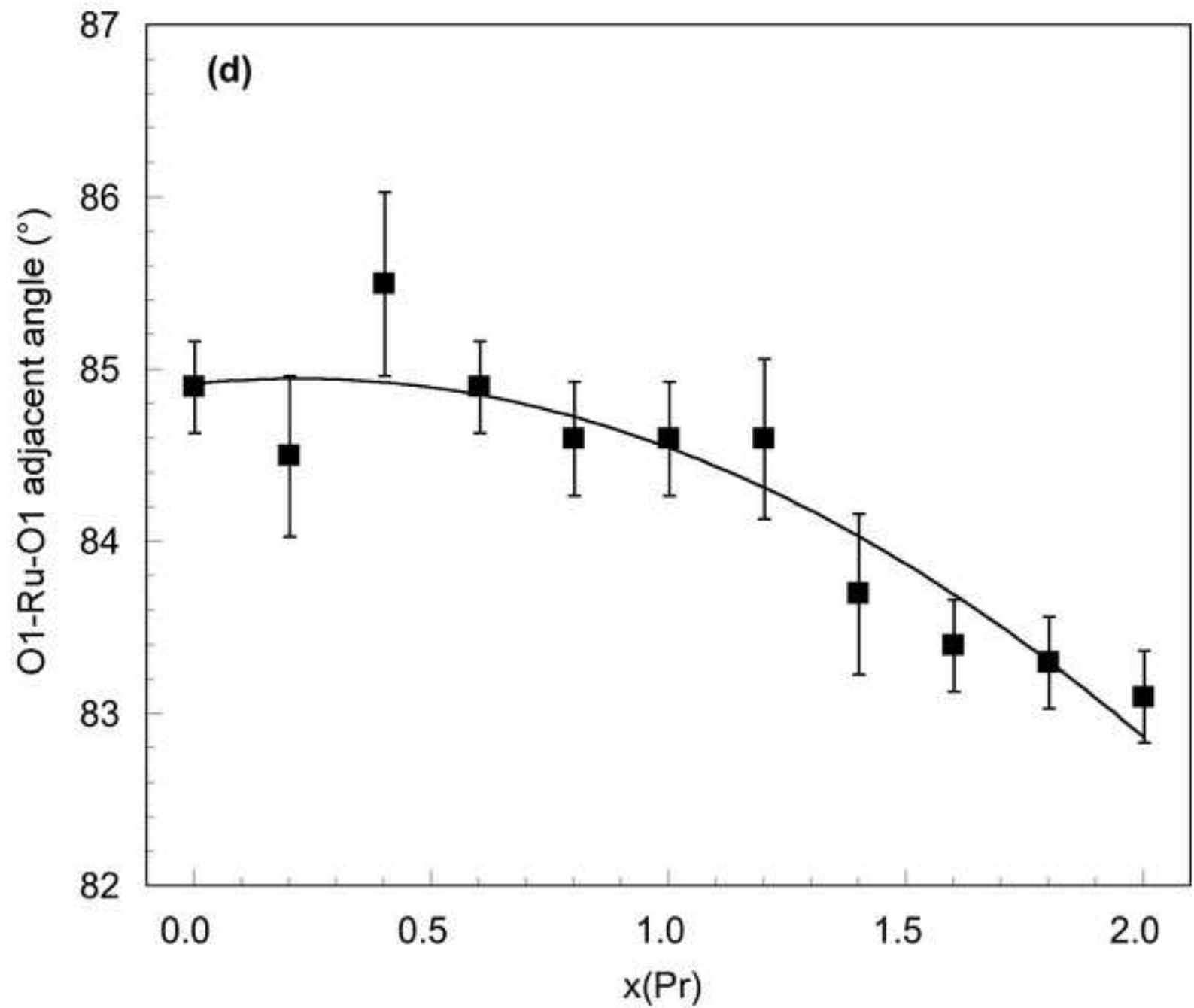

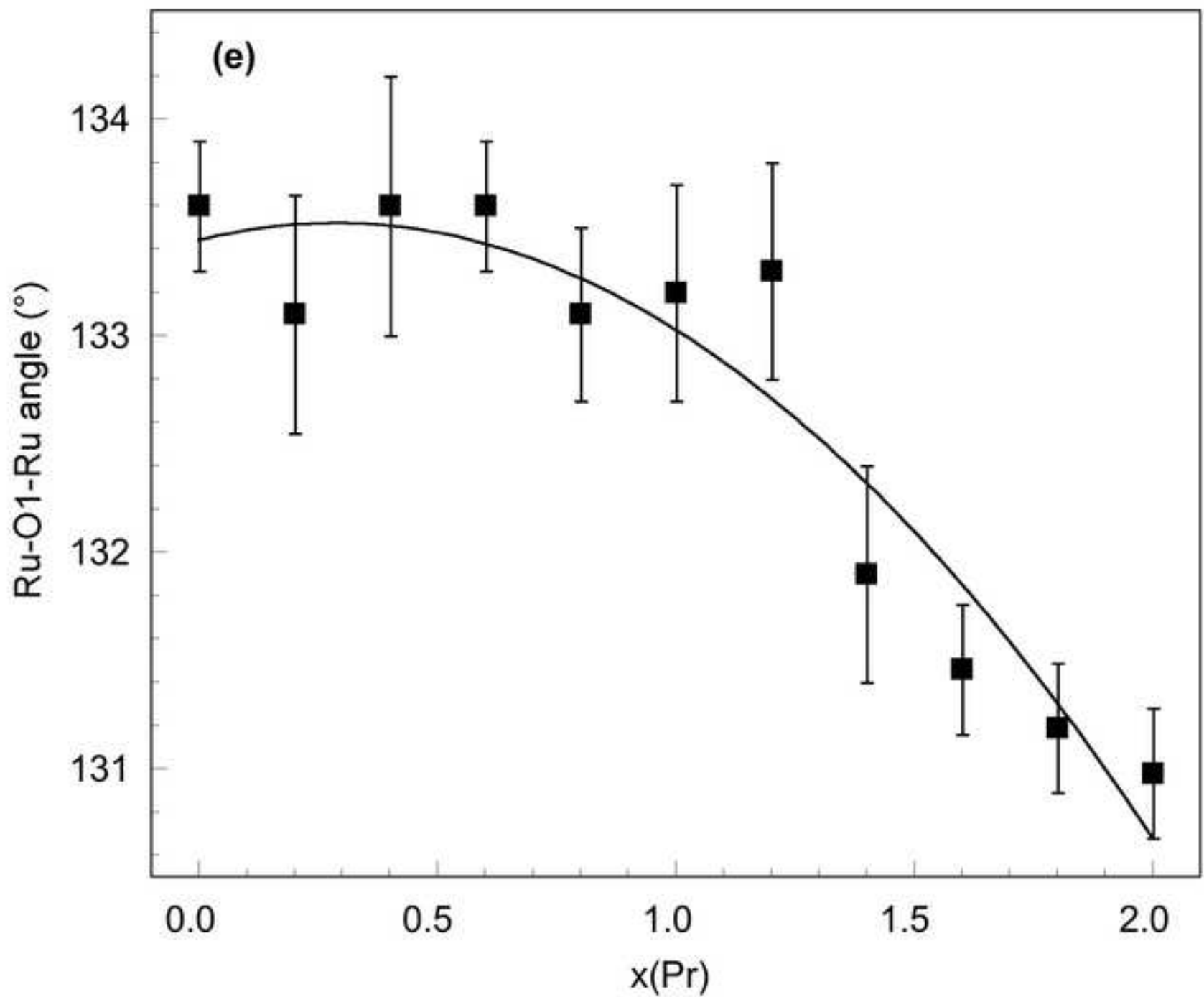

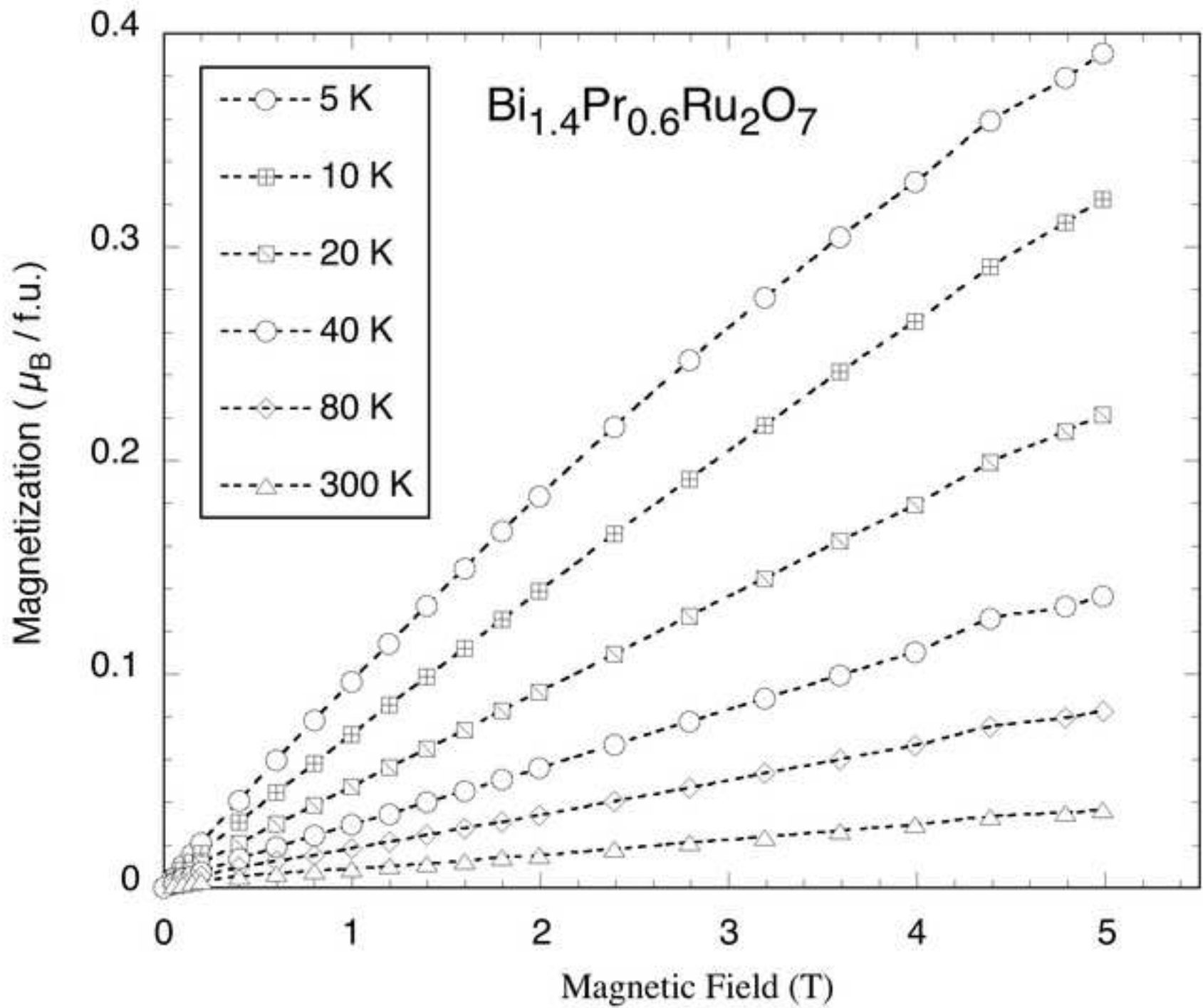

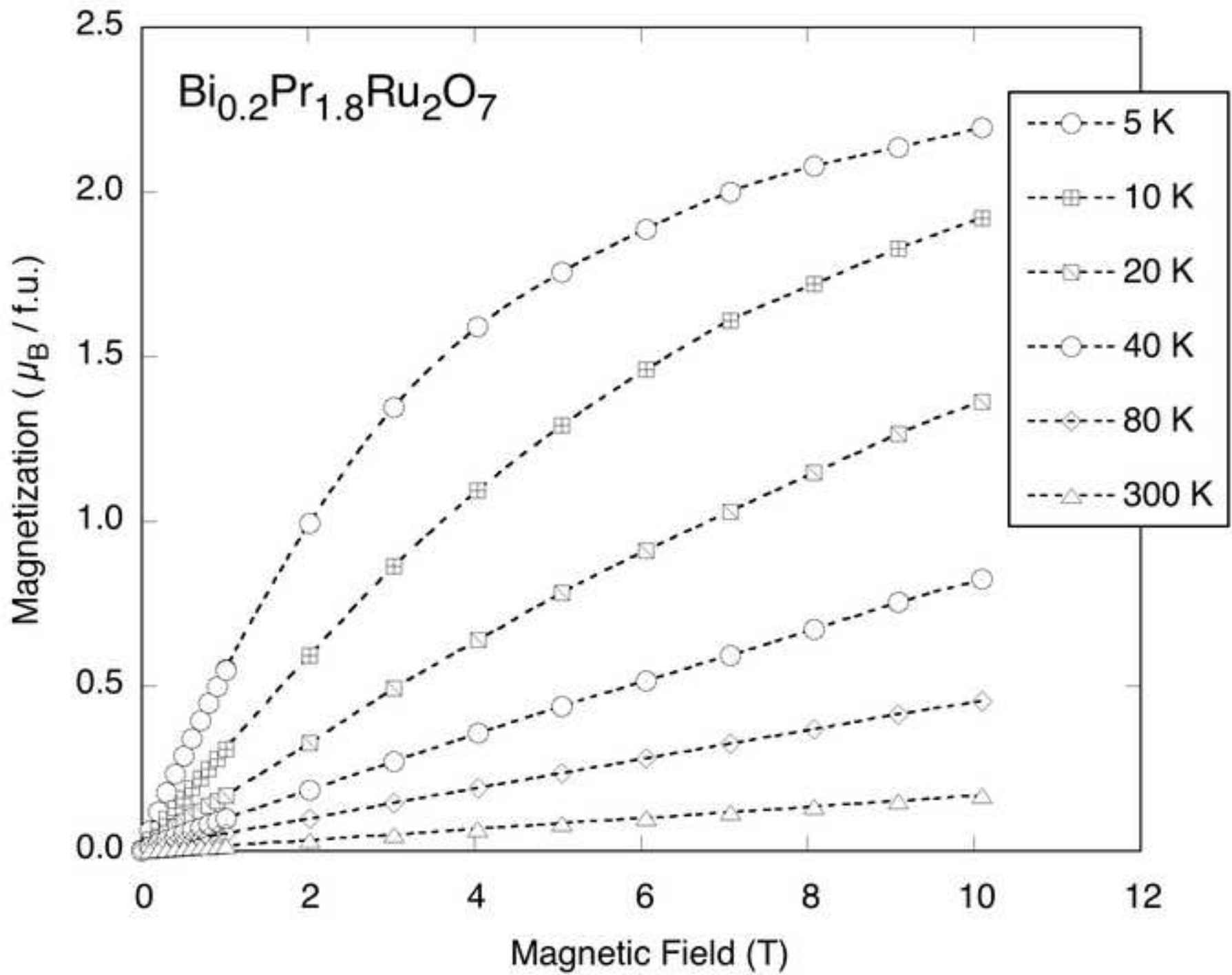

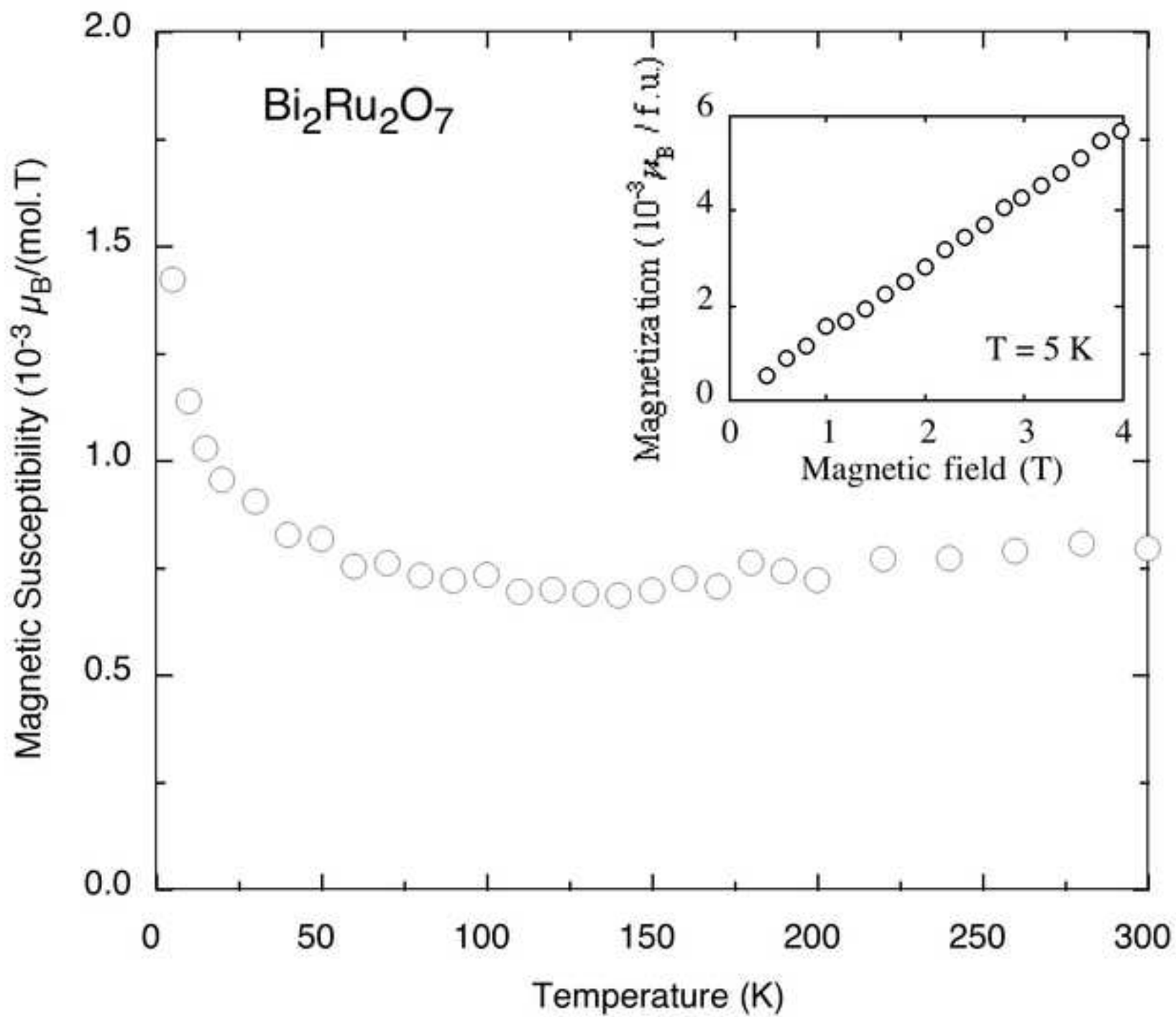

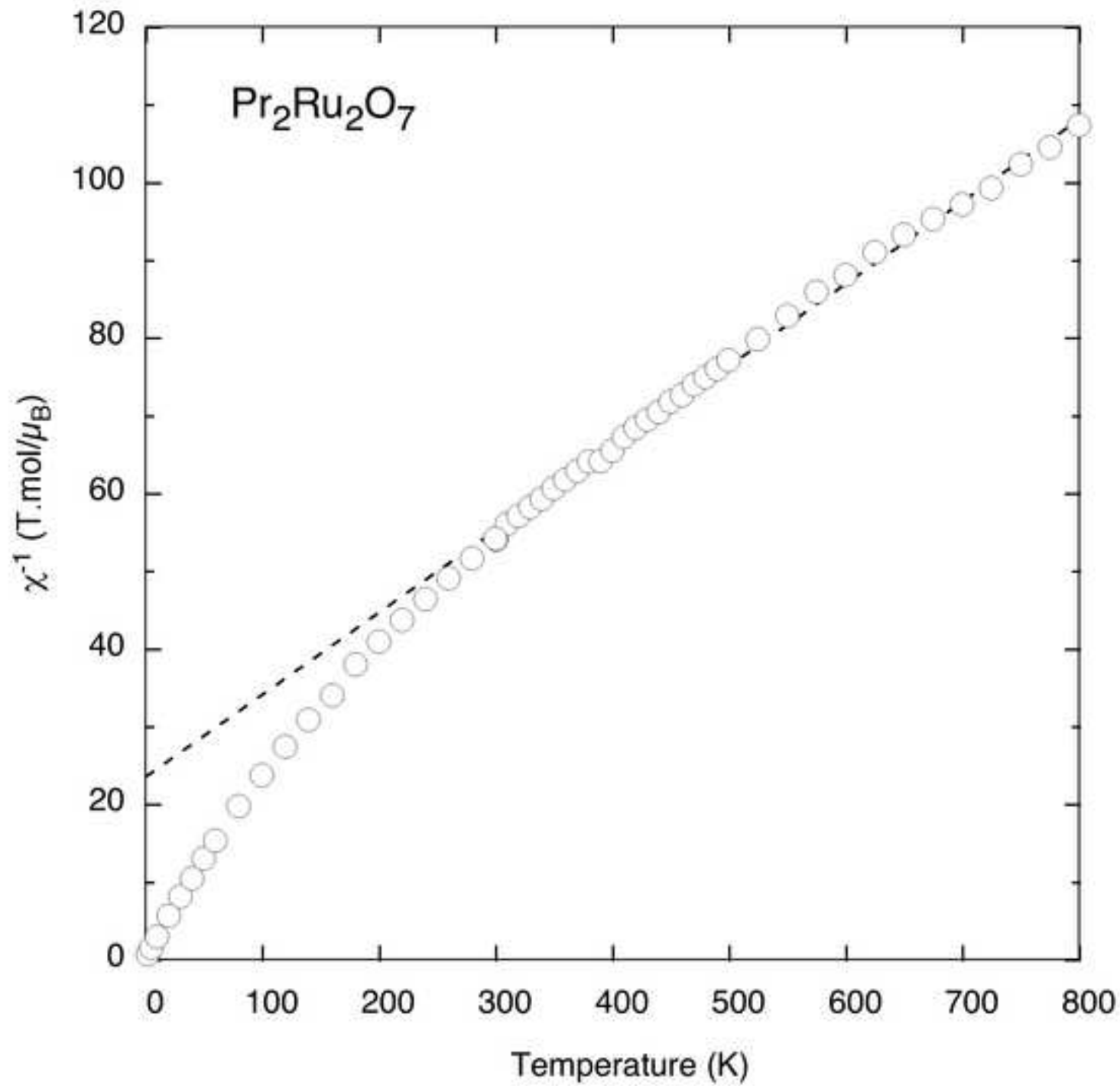